\documentclass[a4paper]{article}

\usepackage{INTERSPEECH2020}
\usepackage{amsmath,graphicx,makecell,url}

\title{}
\title{INT8 Winograd Acceleration for Conv1D Equipped ASR Models Deployed on Mobile Devices}
\name{Yiwu Yao, Yuchao Li, Chengyu Wang, Tianhang Yu, Houjiang Chen, Xiaotang Jiang, Jun Yang, Jun Huang, Wei Lin, Hui Shu, Chengfei Lv}
\address{Alibaba Group, Hangzhou, China}
\email{yiwu.yyw@alibaba-inc.com}

\begin{document}

\maketitle
\begin{abstract}
The intensive computation of Automatic Speech Recognition (ASR) models obstructs them from being deployed on mobile devices.
In this paper, we present a novel quantized Winograd optimization pipeline, which combines the quantization and fast convolution to achieve efficient inference acceleration on mobile devices for ASR models.
To avoid the information loss due to the combination of quantization and Winograd convolution, a Range-Scaled Quantization (RSQ) training method is proposed to expand the quantized numerical range and to distill knowledge from high-precision values.
Moreover, an improved Conv1D equipped DFSMN (ConvDFSMN) model is designed for mobile deployment. 
We conduct extensive experiments on both ConvDFSMN and Wav2letter models. 
Results demonstrate the models can be effectively optimized with the proposed pipeline. Especially, Wav2letter achieves 1.48$\times$ speedup with an approximate $0.07\%$ WER decrease on ARMv7-based mobile devices. 
\end{abstract}
\noindent\textbf{Index Terms}: Range-Scaled Quantization, INT8 Winograd, ConvDFSMN, Mobile deployment

\section{Introduction}
\label{sec:intro}

Recent years has witnessed the great success of deep neural networks in many real-world applications, \emph{e.g.,} image classification~\cite{DBLP:conf/cvpr/HeZRS16}, text translation~\cite{DBLP:journals/corr/BahdanauCB14} and speech recognition~\cite{DBLP:journals/taslp/MohamedDH12}.
While sophisticated neural networks have adequately advanced the performance, the computational workload and the storage requirement are also drastically increased, which obstruct their applications in mobile devices~\cite{DBLP:conf/ICDCS/ICDCSDLM18}.
Especially, Automatic Speech Recognition (ASR) applications require real-time voice interaction, leading to a more challenging field of deploying ASR models on mobile devices~\cite{DBLP:conf/ICASSP/ICASSPJ08}.

In the literature, mainstream ASR models can be classified into feed-forward structures ~\cite{DBLP:conf/ISCA/VDS15,DBLP:journals/corr/CollobertPS16,DBLP:conf/icassp/ZhangLYD18} and auto-regressive structures ~\cite{DBLP:journals/corr/abs-2006-01713,DBLP:conf/ICASSP/LSB18}.
Auto-regressive models (\emph{e.g.} transformer and its variants) exploit beam search to achieve end-to-end speech recognition, but are difficult to accelerate due to the while-loop decoding.
In contrast, feed-forward models avoid the auto-regressive property and produce the outputs in parallel, indicating better inference efficiency.
To enhance the ability of capturing contextual information, several feed-forward acoustic models (\emph{e.g.,} TDNN~\cite{DBLP:conf/ISCA/VDS15} and DFSMN~\cite{DBLP:conf/icassp/ZhangLYD18}) use the one-dimensional convolution (Conv1D) to extract contextual features.
However, the ratio of computation to memory access of the Conv1D is relatively low, which forms the inference performance bottleneck of these ASR models.

Quantization ~\cite{DBLP:journals/corr/abs-1806-08342,DBLP:conf/ICCV/MMTMDF19,DBLP:conf/ICLR/JZSPVK18,DBLP:conf/ICLR/StevenKE20} is a crucial technique to reduce the computation latency and model size when deploying the deep neural networks on mobile devices.
Current mobile inference frameworks (\emph{i.e.,} TFLite~\cite{TFLite17}, MNN~\cite{DBLP:conf/MLSys/Xiaotang20} and NCNN~\cite{NCNN17}) provide 8-bit integers quantization based on matrix multiplication.
Compared to matrix multiplication, Winograd's minimal filtering ~\cite{DBLP:conf/CVPR/LavinA20} reduces the multiplications in convolution by exploiting the correspondence between convolution and scalar multiplication, showing higher efficiency.
Unfortunately, the simple combination of these two techniques will disturb the quantized values and cause a precision overflow~\cite{DBLP:conf/MLSys/JPAM20,DBLP:conf/ICASSP/GLX20}, which obstructs optimizing the model by these two methods simultaneously.

\begin{figure}[t]
  \centering
  \centerline{\includegraphics[width=8.6cm]{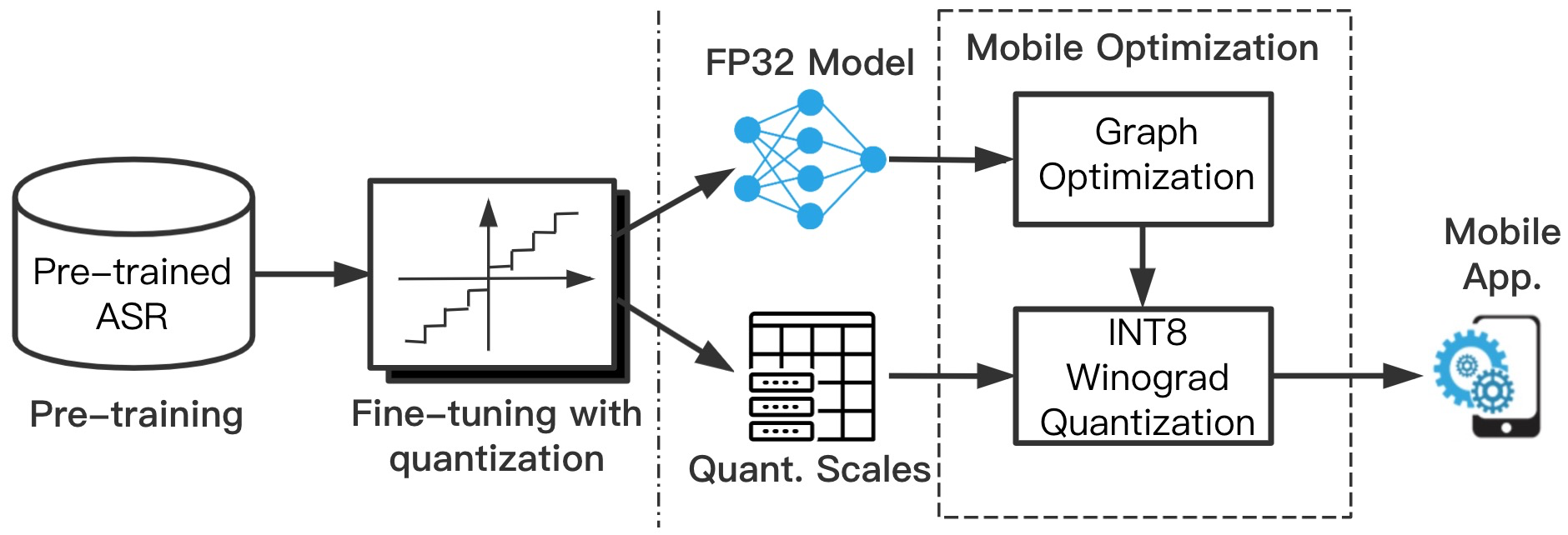}}
  \caption{Proposed optimization pipeline: i) the pre-trained Conv1D equipped ASR model is fine-tuned with RSQ; ii) the re-trained model with quantization scales are sent to mobile optimization flow, where the graph optimization and INT8 Winograd are conducted for mobile deployment.}
\label{fig:workflow}
\end{figure}

In this paper, as shown in Fig.~\ref{fig:workflow}, to accelerate the inference of Conv1D-based ASR models, we propose a novel quantized Winograd method, which integrates the Range-Scaled Quantization (RSQ) training and the low-precision quantized Winograd optimization.
Firstly, the network is fine-tuned with RSQ including integer range scaling and quantization noise minimization, which obtains better quantization scales.
After that, the weights and activations in the network are quantized as lower precision to prevent the addition overflow in quantized Winograd.
Finally, in the Winograd domain, the full integer Hardmard production is executed to improve the efficiency of convolution.

We summarize our four main contributions as follows:

\begin{itemize}
\item We propose the quantization noise loss and the range scaling mechanism during RSQ training, which improves the performance of networks after quantizing.

\item The low-precision quantized Winograd is proposed to speed up the Conv1D operations on mobile devices.

\item A pre-trained, Conv1D-based ASR model is designed based on DFSMN, which has better performance after optimizing by our method.

\item We conduct extensive experiments to verify the effectiveness of our optimization strategies.
\end{itemize}

\section{Range-scaled Winograd Quantization}
\label{sec:RSQ}

In this section, we first show that the quantized range in Winograd quantization is smaller than that of simple quantization because of the addition overflow in Winograd transformation.
To obtain the model with similar performance after Winograd quantization, the range scaling mechanism and quantization noise loss should be jointly applied during fine-tuning.

\subsection{Addition Overflow of Quantized Winograd}
The Conv1D operations (with kernel size $k\ge3$) can be efficiently accelerated with Winograd algorithm~\cite{DBLP:conf/CVPR/LavinA20}.
However, when combining quantization and the Winograd convolution, if we quantize the value before the Winograd convolution, the addition overflow will be suffered during Winograd transformation. 
We introduce the symmetric uniform scheme~\cite{DBLP:journals/corr/abs-1806-08342} to quantize the input activation $d$ and weight $g$ as $t$-bit signed integers.
The quantized Winograd of Conv1D is:
\begin{equation}
S = A^T[(G Q(g)) \odot (B^T Q(d))],
\end{equation}
where $S$ is the output. $A$, $G$ and $B$ are Winograd transformation matrices. $\odot$ represents the Hadamard production.
We assume the elements of the quantized activation $Q(d)$ and weight $Q(g)$ achieve the boundary value $|D_Q|=2^{t-1}-1$ and $|G_Q|=2^{t-1}-1$ (\emph{e.g.,} $D_Q = 127$ when $t=8$ for INT8 Winograd) , then the $F(2,3)$ transformation is derived as:
\begin{equation}
\left[ 
  \begin{matrix}
   1 & 0 & -1 & 0\\
   0 & 1 & 1 & 0\\
   0 & -1 & 1 & 0\\
   0 & -1 & 0 & 1
  \end{matrix}\right]\left[ 
  \begin{matrix}
   D_Q \\
   D_Q \\
   D_Q \\
   D_Q 
  \end{matrix}\right]=\left[ 
  \begin{matrix}
   0 \\
   2D_Q \\
   0 \\
   0 
  \end{matrix}\right]
\end{equation}

\begin{equation}
2\left[ 
  \begin{matrix}
   1 & 0 & 0\\
   1/2 & 1/2 & 1/2 \\
   1/2 & -1/2 & 1/2 \\
   0 & 0 & 1
  \end{matrix}\right]\left[ 
  \begin{matrix}
   G_Q \\
   G_Q \\
   G_Q 
  \end{matrix}\right]=\left[ 
  \begin{matrix}
   2G_Q \\
   3G_Q \\
   G_Q \\
   2G_Q
  \end{matrix}\right]
\end{equation}
It shows that the outputs of Winograd transformation for input activation $2D_Q$ and weight $3G_Q$ are both out of the quantized range $[1-2^{t-1}, 2^{t-1}-1]$.
To avoid the addition overflow, $Q(d)$ and $Q(g)$ should be in the smaller range $[\frac{1-2^{t-1}}{2}, \frac{2^{t-1}-1}{2}]$ and $[\frac{1-2^{t-1}}{3}, \frac{2^{t-1}-1}{3}]$, respectively.

\subsection{Range Scaling Mechanism}

\begin{figure}
\centering
\includegraphics[width=\columnwidth]{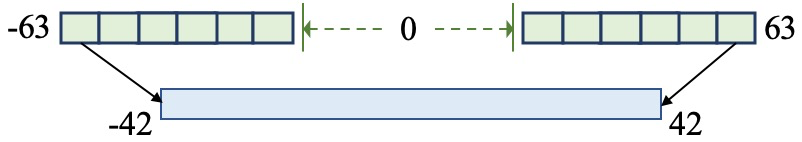}
\caption{Integer range scaling: the 7-bit integer-range [-63, 63] is scaled to [-42, 42] for weight quantization of Conv1D with INT8 Winograd.}
\label{fig:range_scale}
\end{figure}

According to the anti-overflow conditions described previously, when $t=8$, the range of $Q(g)$ is $[-42, 42]$, which between the range of 7-bit and 6-bit.
It makes the weight $g$ only can be quantized to 6-bit (whose quantized range is $[-31, 31]$) for general quantization tools, which cause the larger information loss.
Therefore, we propose the range scaling mechanism to enlarge the integer-range appropriately.

Denoting a wide integer range as $[-T, T]$, a scaling factor $\alpha$ is specifically introduced to obtain the scaled integer-range $[-T_s, T_s]=[-T/\alpha, T/\alpha]$. 
As shown in Fig.~\ref{fig:range_scale}, the weights can be quantized in the scaled 7-bit interval $[-42,42]$ by setting $\alpha=1.5$, thus the numerical representation becomes richer than the 6-bit integer-range $[-31, 31]$. Additionally, $\alpha=1.0$ for 7-bit quantization of activations.

\subsection{Quantization Noise Loss}

Usually, large-scale, pre-trained ASR models should be fine-tuned on specific domains to adapt to downstream tasks.
To simplify the process and ensure the prediction accuracy after quantization, the Quantization-Aware Training (QAT) ~\cite{DBLP:conf/ICLR/JZSPVK18,DBLP:conf/ICLR/StevenKE20} is introduced for fine-tuning.

Here, the Learned Step-size Quantization (LSQ)~\cite{DBLP:conf/ICLR/StevenKE20} is adopted as the basic QAT method.
Based on LSQ, the tensor $v$ (weight or activation) is fake quantized with the quantization scale $s$ and scaled integer range $[-T_s, T_s]$ as below:

\begin{equation}
\label{eq:quant}
Q(v) = s \cdot round(clip(v/s, -T_s, T_s))
\end{equation}

The gradients can be formulated as follows:
\begin{equation}
\label{eq:quant_grad}
\dfrac{\partial Q}{\partial v} = \left\{
             \begin{array}{lr}
             1, & (-T_s \leq v/s \leq T_s) \\
             0, & (otherwise)  
             \end{array} \right.
\end{equation}
\begin{equation}
\label{eq:scale_grad}
\dfrac{\partial Q}{\partial s} = \left\{
             \begin{array}{lr}
             -T_s, & (v/s < -T_s) \\
             -v/s+round(v/s), & (-T_s \leq v/s \leq T_s) \\
             T_s, & (v/s > T_s)  
             \end{array} \right.
\end{equation}
To enhance the effectiveness of LSQ, we introduce the mean squared error between $q(v)$ and $v$ as an auxiliary quantization noise loss to distill knowledge from high-precison values, which makes the quantization values close to original values:
\begin{equation}
\label{eq:mse_loss}
L_q=MSE(Q(v),v)=\frac {1}{N}(Q(v)-v)^2
\end{equation}

The gradients of $L_q$ w.r.t $v$ and $L_q$ w.r.t $s$ are derived as:
\begin{equation}
\label{eq:mse_quant_grad}
\frac{\partial L_Q}{\partial v}=\frac {2}{N}(Q(v)-v)(\frac{\partial Q}{\partial v}-1)
\end{equation}
\begin{equation}
\label{eq:mse_scale_grad}
\frac{\partial L_Q}{\partial s}=\frac {2}{N}(Q(v)-v)\frac{\partial Q}{\partial s}
\end{equation}

Combining Eq.~\ref{eq:quant_grad} and Eq.~\ref{eq:mse_quant_grad}, the gradient is less than 0 when $v/s < -T_s$, and is greater than 0 when $v/s > T_s$.
Thus, the values of tensor $v$ outside $[-sT_s, sT_s]$ will be updated towards a more concentrated distribution, friendly to quantization.
Additionally, under the consideration of Eq.~\ref{eq:scale_grad} and Eq.~\ref{eq:mse_scale_grad}, the scale $s$ will be updated to achieve the suitable quantization resolution for the tensor values inside or outside $[-sT_s, sT_s]$.

\section{Mobile Optimization with INT8 Winograd}
\label{sec:MNN}

\begin{figure}
\centering
\includegraphics[width=\columnwidth]{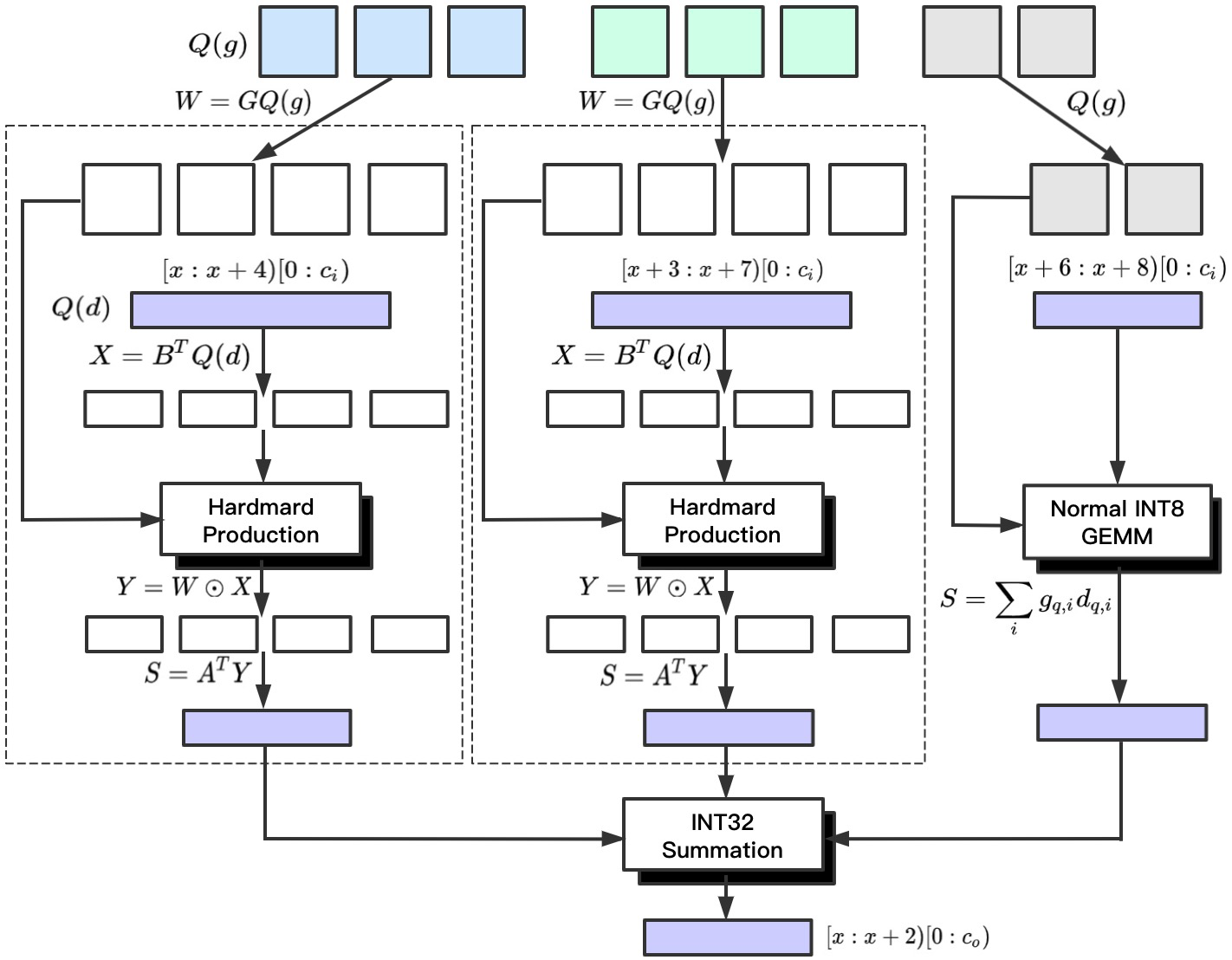}
\caption{INT8 Winograd Operator of Conv1D (e.g., kernel size $k=8$): the Conv1D is split as the basic $F(2,3)$ Winograd convolution and remaining normal INT8 GEMM.}
\label{fig:winograd_op}
\end{figure}

The efficient mobile inference framework MNN~\cite{DBLP:conf/MLSys/Xiaotang20} is employed to deploy the ASR models on mobile devices for real-time applications. To further speed up the Conv1D equipped ASR models, the efficient INT8 Winograd operator is designed on ARMv7-based devices.

When Conv1D with kernel size ($k\ge3$) and stride ($s=1$) is quantized as INT8 Winograd representation, where the activation and weight are expressed as 7-bit integers in the non-Winograd domain, the INT8 Winograd operator will be mapped for mobile inference. As depicted in Fig.~\ref{fig:winograd_op}, the designed INT8 Winograd operator is split into several $F(2,3)$ formations and the remaining normal INT8 GEMM (General Matrix Multiplication). In each flow of the $F(2,3)$, the sub-sampled 7-bit ($3\times1$) weight and ($4\times1$) activation are firstly transformed to Winograd domain. The fully INT8 Hardmard production is conducted for highly efficient execution of Conv1D. Finally, the results of each Winograd flow and normal GEMM flow are summed as the convolution output. 

Compared with the normal INT8 GEMM realization, the theoretical speedup of INT8 Winograd Conv1D can be expressed with kernel size ($k\ge3$) as below:
\begin{equation}
\label{eq:speedup}
speedup=\frac{2k}{4\lfloor k/3 \rfloor+2(k\%3)}
\end{equation}
The maximum speedup is $1.5$ with $k=3m,m\in {N}^+$.

\section{Conv1D Equipped ASR Model}
\label{sec:asr}

To fully leverage the INT8 Winograd optimization technique, a simplified ASR model is designed for mobile applications, namely the Conv1D Equipped Deep Feed-Forward Sequential Memory Network (ConvDFSMN). 

\subsection{Model Architecture}

\begin{figure}
\centering
\includegraphics[width=\columnwidth]{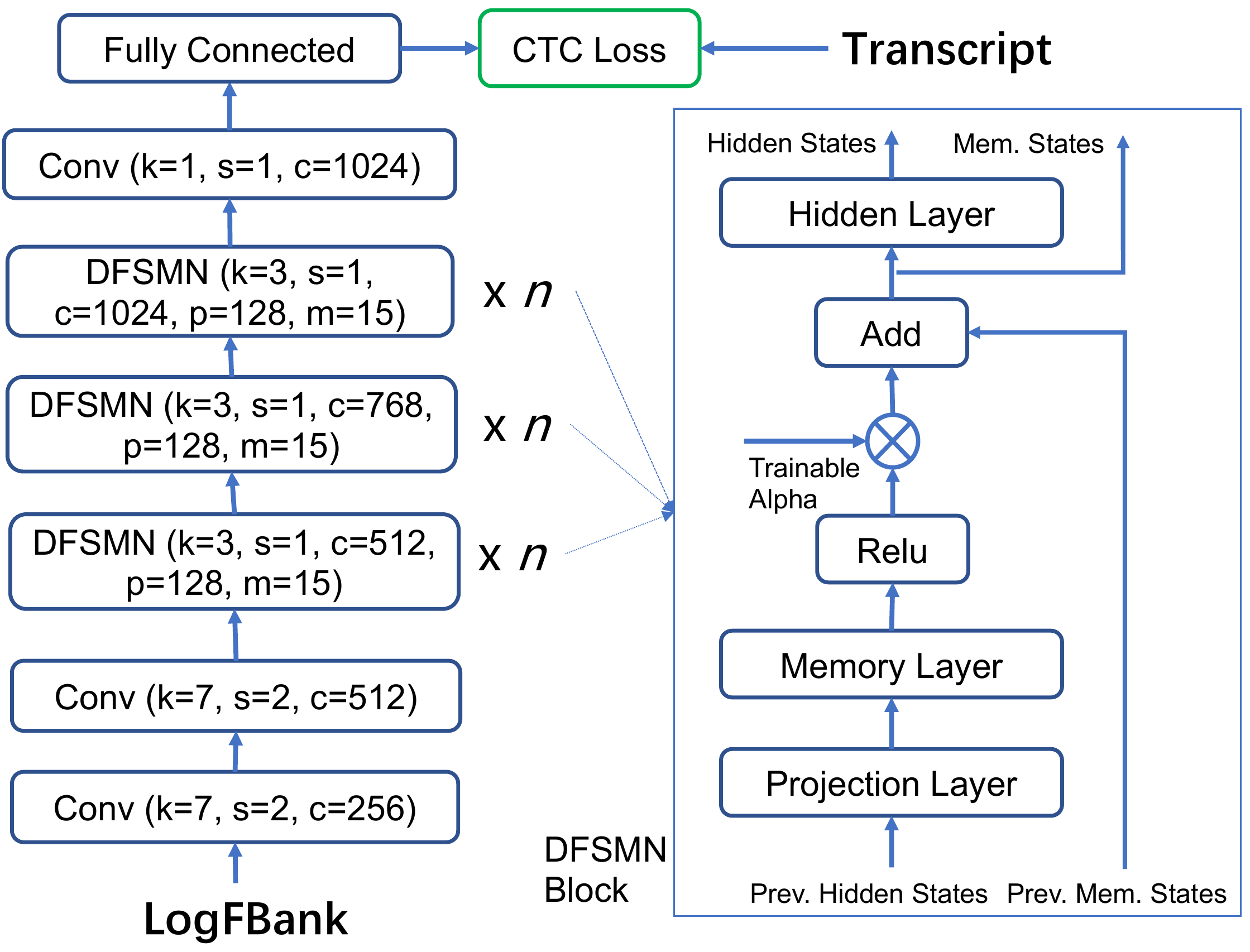}
\caption{ConvDFSMN architecture. Notations: $k$: kernel size, $s$: stride, $c$: hidden size, $p$: project size, $m$: memory size. The DFSMN block repetition factor is set as: $n$=2 or 4.}\label{fig:model}
\end{figure}

The model architecture of ConvDFSMN is illustrated in Fig.~\ref{fig:model}. It has a highly compact structure extended from Wav2letter~\cite{DBLP:journals/corr/CollobertPS16} and DFSMN~\cite{DBLP:conf/icassp/ZhangLYD18}, with the CTC (Connectionist Temporal Classification) loss~\cite{DBLP:conf/icml/GravesFGS06} as its main loss function. 

At the bottom of ConvDFSMN, Conv1D layers~\cite{DBLP:journals/corr/CollobertPS16} are applied to the input features, allowing the model to ``see'' local contexts of waves. After that, a few Conv1D equipped DFSMN blocks~\cite{DBLP:conf/icassp/ZhangLYD18} (with $n=2$ or 4 in Fig.~\ref{fig:model}) are employed to model the recurrent characteristics of human speeches. The memory layers inside DFSMN is used to store the memory states of the input feature sequence, significantly reducing computational complexities compared to RNN-based models. As Conv1D can ``summarize'' the contextual features locally, the need for stacked DFSMN blocks to obtain an accurate and concise network architecture is decreased.

\subsection{Large-scale Model Pre-training}
The training of ConvDFSMN is non-trivial for the purpose of generating a sufficiently small model with high accuracy. The highly limited parameter space of ConvDFSMN makes the model challenging to reach the global optima.
In this work, we employ the weakly supervised learning technique~\cite{DBLP:journals/corr/abs-2008-01300} for pre-training, with human speech-transcript pairs extracted from massive video data. The total duration is approximately 8,000 hours. The weakly supervised pre-training module has been integrated into the EasyASR platform~\cite{DBLP:journals/corr/abs-2009-06487} to support large-scale distributed training. After pre-training, the model is further fine-tuned on domain-specific datasets with RSQ.

\section{Experiments}
In this section, we evaluate the proposed pipeline in detail.
\label{sec:experiments}

\subsection{Experimental Settings}

We introduce the ConvDFSMN-b (-base) with $n=4$ and ConvDFSMN-s (-small) with $n=2$ to verify the effectiveness of our method.
In addition, the simplified Wave2letter model~\cite{DBLP:journals/corr/CollobertPS16} with almost the equivalent parameter size to ConvDFSMN-b is designed for comparison.
Details about these three models are shown in Table~\ref{tab:model_des}. 

\begin{table}
\begin{footnotesize}
\begin{tabular}{llllll} 
\hline
\bf Model  & \bf Parameters (M)    &  \bf FLOPs (G)  & \bf Latency (ms) \\ 
\hline
ConvDFSMN-b  &   16.20 (6.52)  & 2.45 (0.98)  & 63.5 (28.1)  \\
\hline
ConvDFSMN-s  &   12.37 (6.52)  & 1.87 (0.98) & 53.4 (28.5)  \\
\hline
Wav2letter &   17.32 (6.52)  &  2.62 (0.98) &  87.2 (28.5)      \\
\hline
\end{tabular}
\caption{Description of three FP32 models. The FLOPs and latency are tested with input sequence length equalling to 600. Contents in $()$ are attributed to the classifier layers dominating the size and computation of involved models.}
\label{tab:model_des}
\end{footnotesize}
\end{table}

\begin{table}
\begin{footnotesize}
\begin{tabular}{llllll} 
\hline
\bf Model   & \bf Quant. Type &  \bf Method &  \bf WER   & \bf Latency  \\ 
\hline
ConvDFSMN-b$^*$ & FP32 Wino.     &   -    & 8.80\%  & 63.5ms  \\
                & INT8 GEMM      &   PTQ  & 8.89\%  & 59.1ms  \\
                & INT8 Wino.     &   PTQ  & 8.94\%  & 56.9ms  \\
\hline
ConvDFSMN-b     & INT8 Wino.  &   RSQ\dag  &  8.76\% & 56.9ms  \\
                & INT8 Wino.  &   RSQ      &  8.73\% &  -    \\
\hline
ConvDFSMN-s$^*$ & FP32 Wino.      &   -    & 10.54\%  & 53.4ms  \\
                & INT8 GEMM     &  PTQ   & 10.62\%  & 50.7ms  \\
                & INT8 Wino.      &  PTQ   & 10.68\%  & 48.6ms  \\
\hline
ConvDFSMN-s     & INT8 Wino.   &  RSQ\dag  & 10.53\% & 48.6ms   \\
                & INT8 Wino.   &  RSQ      & 10.52\%  &  -      \\
\hline
Wav2letter$^*$  & FP32 Wino.   &   -    &  13.78\%  & 87.2ms \\
                & INT8 GEMM  &  PTQ   &  13.84\% & 63.2ms      \\
                & INT8 Wino.   &  PTQ   &  13.87\% & 59.1ms     \\
\hline
Wav2letter      & INT8 Wino.   &  RSQ\dag   & 13.73\%  & 59.1ms      \\
                & INT8 Wino.   &  RSQ   & 13.71\%  & -     \\
\hline
\end{tabular}
\caption{Comparison of models on Aishell-1 and latency. Models with (*) are fine-tuned without RSQ. RSQ with (\dag) means the MSE loss is not employed. The classifier sensitive to quantization is not quantized for all models, and the Conv1D with $k=1$ or $s>1$ is quantized as normal INT8.}
\label{tab:model_acc}
\end{footnotesize}
\end{table}

All the ASR models are pre-trained on EasyASR~\cite{DBLP:journals/corr/abs-2009-06487}, then fine-tuned and tested on Aishell-1~\cite{DBLP:conf/O-COCOSDA/HBu17}.
We finetune the network for 3000 steps with the mini-batch size of 128 and introduce the ploy decay of learning rate with initial $lr=0.005$.
The MSE loss in Eq.~\ref{eq:mse_loss} is multiplied with $\beta=0.25$.
KL algorithm referred in TensorRT~\cite{TRT20} is applied for conducting the Post-Training Quantization (PTQ), and initializing the quantization scales of RSQ.
The inference latency of proposed models is profiled on K20 Pro mobile phone with ARMv7 ISA (4 threads) and batch size equal to 1.

\subsection{Experimental Results}

As demonstrated in Table~\ref{tab:model_acc}, the performance of ConvDFSMN is better than Wav2letter, which indicates the superiority of ConvDFSMN.
Even the ConvDFSMN-s (with $12.37M$ parameters) achieves $10.5\%$ WER on Aishell, which is lower than the result of Wav2letter (with $17.32M$ parameters).

The quantized models fine-tuned with RSQ achieve better performance than models quantized with PTQ, even the Conv1D is represented as INT8 Winograd.
The experiments in Table~\ref{tab:model_acc} shows that the MSE loss can success distill knowledge from high-precision value, which make network achieve the better performace after quantization.
The advantages of integrating the RSQ into fine-tuning are two-folds: i) the quantization flow is simplified by omitting the extra PTQ with the calibration-set; and ii) RSQ can act as the regularizer to improve the network performance after quantization.

In terms of inference latency, compared with the normal INT8 GEMM, the networks gain further speedup with INT8 Winograd optimization on mobile devices.
Table~\ref{tab:conv1d_latency} shows the runtime latency of Conv1D used in ConvDFSMN and Wav2letter, where the INT8 Winograd shows positive optimization for all used kernel sizes.
However, the actual speedup of INT8 Winograd is constrained to the ratio of computation to memory access.
So that the Conv1D with 15$\times$1 kernel size in ConvDFSMN only achieves 1.1$\times$ speedup, and the similar Conv1D in Wav2letter can achieve $1.3\times$ speedup, of which the theoretical speedup is $1.5\times$.

\begin{table}
\begin{footnotesize}
\begin{tabular}{llllll} 
\hline
\bf Model  & \bf Kernel Size & \bf  GEMM (ms) & \bf Wino. (ms) \\ 
\hline
        & (3,1,128, 512)  & \makecell[c]{0.601} & 0.536 (1.12$\times$) \\
ConvDFSMN & (3,1,128, 768)  & \makecell[c]{0.898} & 0.788 (1.14$\times$) \\
          & (3,1,128, 1024) & \makecell[c]{0.881} & 0.710 (1.24$\times$) \\
          & (15,1,128, 128) & \makecell[c]{0.540} & 0.491 (1.10$\times$)\\
\hline
          & (9, 1, 256, 512) & \makecell[c]{3.42} & 2.75 (1.24$\times$) \\
Wav2letter & (13, 1, 512, 512) & \makecell[c]{9.70} & 8.27 (1.17$\times$)\\
           & (15, 1, 512, 512) & \makecell[c]{9.29} & 7.13 (1.30$\times$)\\
\hline
\end{tabular}
\caption{Latency of the INT8 Conv1D. The input sequence length of listed Conv1D operations is 150. Both INT8 GEMM and INT8 Winograd operators are profiled for comparison.}
\label{tab:conv1d_latency}
\end{footnotesize}
\end{table}

\section{Conclusion}
\label{sec:conclusion}

To derive the motivation of real-time voice interaction on mobile devices, an quantized Winograd pipeline with range-scaled quantization (RSQ) training and INT8 Winograd realization is presented in this paper.
The designed Conv1D equipped DFSMN (ConvDFSMN) can be effectively quantized and efficiently accelerated with the proposed method, indicating a significant potential for industrial applications.


\end{document}